\documentclass[5p]{elsarticle}
\usepackage[utf8]{inputenc}
\usepackage{lineno,hyperref}
\usepackage{xcolor}
\usepackage{makecell}
\usepackage{amssymb}
\usepackage{amsmath}
\usepackage{stfloats}
\usepackage{color,soul}
\modulolinenumbers[5]

\journal{Journal of \LaTeX\ Templates}

\begin{document}

\begin{frontmatter}

\title{Molecular characterization of macroscopic aerogels of single-walled carbon nanotubes }


\author[IMDEA Materials Institute]{Bel\'{e}n Alem\'{a}n}

\author[IMDEA Materials Institute]{Juan J.  Vilatela \corref{correspondingauthor}}

\ead{juanjose.vilatela@imdea.org}

\address[IMDEA Materials Institute]{ IMDEA Materials Institute. Eric Kandel, 2, Tecnogetafe, 28906, Getafe, Madrid (Spain)}

\cortext[mycorrespondingauthor]{Corresponding author}


\begin{abstract}
Single-walled carbon nanotubes (SWCNT) can be assembled into various macroscopic
architectures, most
notably continuous fibers and films, produced currently on a kilometer-per-day scale by
floating catalyst
chemical vapor deposition and spinning from an aerogel of CNTs. An attractive challenge is to produce continuous
fibers with
controlled molecular structure with respect to the diameter, chiral
angle and
ultimately \textit{(n,m)} indices of the constituent SWCNT “molecules”. This work
presents an extensive
Raman spectroscopy and high-resolution transmission electron microscopy study of
SWCNT aerogels produced
by the direct spinning method. By retaining the open structure of the SWCNT aerogel, we
reveal the
presence of both semiconducting and metallic SWCNTs and determine a full distribution of
families of SWCNT
grouped by optical transitions. The resulting distribution matches the chiral angle distribution obtained by electron microscopy and electron diffraction. The effect of SWCNT
bundling on the Raman spectra, such as the G$^{-}$ line shape due to plasmons activated
in the far-infrared
and semiconductor quenching, are also discussed. By avoiding full aggregation of the
aerogel and applying
the methodology introduced, rapid screening of molecular features can be achieved in large
samples, making this protocol a useful analysis tool for engineered SWCNT
fibers and related systems.
\end{abstract}

\begin{keyword}
\texttt{SWCNT, chiral angle, fiber, metallic, semiconducting, Raman}
\end{keyword}

\end{frontmatter}


\section{Introduction}
The two main challenges to exploit the exceptional mechanical, electrical, thermal and
optical properties of carbon nanotubes (CNT) on a macroscopic scale, are the synthesis of building blocks with molecular control, and their organized assembly into larger structures \cite{liu_macroscopic_2011,Guadalupe}. Macroscopic fibers of aligned CNT are a particularly attractive format that naturally maximizes the contribution from longitudinal properties of CNTs. Extensive efforts to improve synthetic methods and fiber spinning processes have led to continuous CNT fibers with tensile mechanical properties on par with carbon fibers \cite{koziol2007high,Lee,NatMat}, thermal conductivity above copper \cite{GSPANN,PasqualiScience} and mass-normalized electrical conductivity above that of metals \cite{PasqualiScience,PasqualiAdv}. These properties are encouraging, but still represent a small fraction of those of the constituent CNTs and thus leave large room for improvement.

Key for the realization of CNT fibers with superior properties to those currently available is the control over the morphology of the constituent CNT in terms of number of layers, diameter and chiral angle. These parameters affect the packing of CNTs in bundles and the stress transfer between them \cite{model}, hence the accepted view in the field that few-layer, large-diameter CNTs lead to superior tensile properties. Better packing of CNTs is also likely to favour charge transfer, but fiber longitudinal electrical conductivity depends very strongly on the content of metallic versus semiconducting CNTs. Fibers of single-walled carbon nanotubes (SWCNT) of specific metallicity are yet to be produced, but work on transparent conductors, for example, shows a conductivity enhancement of around $5.6$ for samples of metallic SWCNTs compared to samples with unsorted SWCNTs \cite{Hersam}. A similar enhancement applied to existing CNT fibers currently made up of mixtures of SWCNTs would lead to a longitudinal conductivity of around 40·10$^{6}$ S/m, or about 4 times higher than copper on a mass basis. 

There are two hypothetical routes towards producing a continuous, macroscopic fiber of single chirality SWCNTs: wet spinning from a dispersion of pre-synthesized and sorted SWCNTs \cite{Vigolo,Ericson}, or by direct spinning of a fiber from an aerogel of chirallity-controlled SWCNTs synthesized by floating catalyst chemical vapor deposition (FCCVD). The direct FCCVD spinning method offers established control over the CNT number of layers \cite{reguero_controlling_2014,SulphurLee}, diameter \cite{Se}, and degree of alignment. But the determination of further molecular features, particularly the fraction of metallic SWCNTs has proven elusive. This is partly a consequence of the fundamental difficulty in characterising individual molecular features (i.e. chiral angle, diameter, etc) in a system with around $10^{8}$ SWCNTs per cross section and $10^{11}$ per metre of fiber in a highly aggregated format. Resonant Raman spectroscopy has been an ideal method to probe (relatively) large sample volumes while resolving fine molecular futures. Yet, studies on SWCNT fibers produced by FCCVD typically extract limited information beyond identification of CNT number of layers and D/G ratio \cite{sundaram_continuous_2011, liu_diameter-selective_2008, barnard_role_2016,ding_highly_2017}.  An approximate chiral angle distribution can be obtained from extensive electron diffraction measurements on SWCNT bundles in a CNT fiber \cite{aleman_inherent_2016}, but the method should be complemented by other techniques, especially those enabling faster characterization.  

In this work, we introduce a method for the screening of fine molecular features of CNT fibers made of SWCNTs by coupling high-resolution electron microscopy analysis and Raman spectroscopy mapping of individual SWCNT bundles of macroscopic aerogels. This strategy unveils a large abundance of previously masked semiconducting SWCNTs, and ultimately leads to the determination of the full distribution of \textit{(n,m)} indices for the SWCNT fiber samples and a relative ratio of metallic to semiconducting tubes. It is intended to accelerate improvement of SWCNT fibers by enabling determination of SWCNT polydispersity (electronic and geometric) in continuous fibers, and ultimately assisting in the quest towards producing single chiral angle SWCNT fibers.

\section{Experimental section}
\subsection{Sample preparation}
SWCNT macroscopic material was synthesized by direct spinning of a SWCNT aerogel produced by FCCVD, using butanol as carbon source, ferrocene as Fe catalyst and thiophene as sulfur catalyst promoter. Synthesis of predominantly SWCNTs was achieved by adjusting the precursor weight fraction at 99.1:0.8:0.1 \cite{reguero_controlling_2014}. Two sample formats were produced: SWCNT fiber and an aerogel film (Fig. \ref{furnace}(a-b)).  The low-density aerogel films correspond to samples of SWCNT aerogel that were directly collected at the exit of the reactor. A SWCNT fiber is produced by densifying the free-standing SWCNT aerogel through capillary forces exerted by a volatile liquid applied at the exit of the FCCVD reactor.  These two types of samples differ in their degree of aggregation, but they are otherwise chemically identical. 

\subsection{Characterization}
\footnotesize
Morphological characterization of the aerogel film was carried out using a dual
beam with field-emission scanning electron microscope (FEGSEM) Helios NanoLab 600i (FEI)
at 10 keV. For extracting the SWCNT and SWCNT bundle diameter distributions, over 100 high-resolution transmission electron  microscope (HRTEM) images were obtained using a Talos F200X (FEI) TEM. SWCNT and bundle diameters were determined by image analysis of these micrographs using Image J software (examples and processing in ESI).
Samples for TEM analysis were directly deposited onto TEM grids at the exit of the reactor, thus avoiding degradation by sonication or other undesired effects induced during dispersion in solvents. 
\footnotesize
Raman spectroscopy was performed with a Renishaw PLC with 532 nm wavelength laser and a Bruker Senterra equipped with 532 nm, 633 nm and 785 nm waveleght lasers. The measurements of directly performed on SWCNT bundles were carried out with 100x/0.85 and 50x/0.75 Leica microscope objectives using a low power configuration in order to keep the laser focused and avoid heating effects. Analizing the SWCNT aerogel under these conditions enables the characterization of individual bundles (ESI). Indeed, the largest lateral distance between bundles in the film is around 1.5-2 $\mu$m, slightly larger that the calculated laser beam spot of 0.8-1 $\mu$m. However, not that all spectra collected when mapping the sample correspond to individual bundles (Examples of spectra for multiple bundles, as well as an example showing the loss of semiconductor features when using a large laser spot size are included in ESI).

\section{Results and discussion}

\subsection{Exposing molecular features in SWCNT aerogels}

\begin{figure}[h]
\centering
  \includegraphics[width=8.5cm]{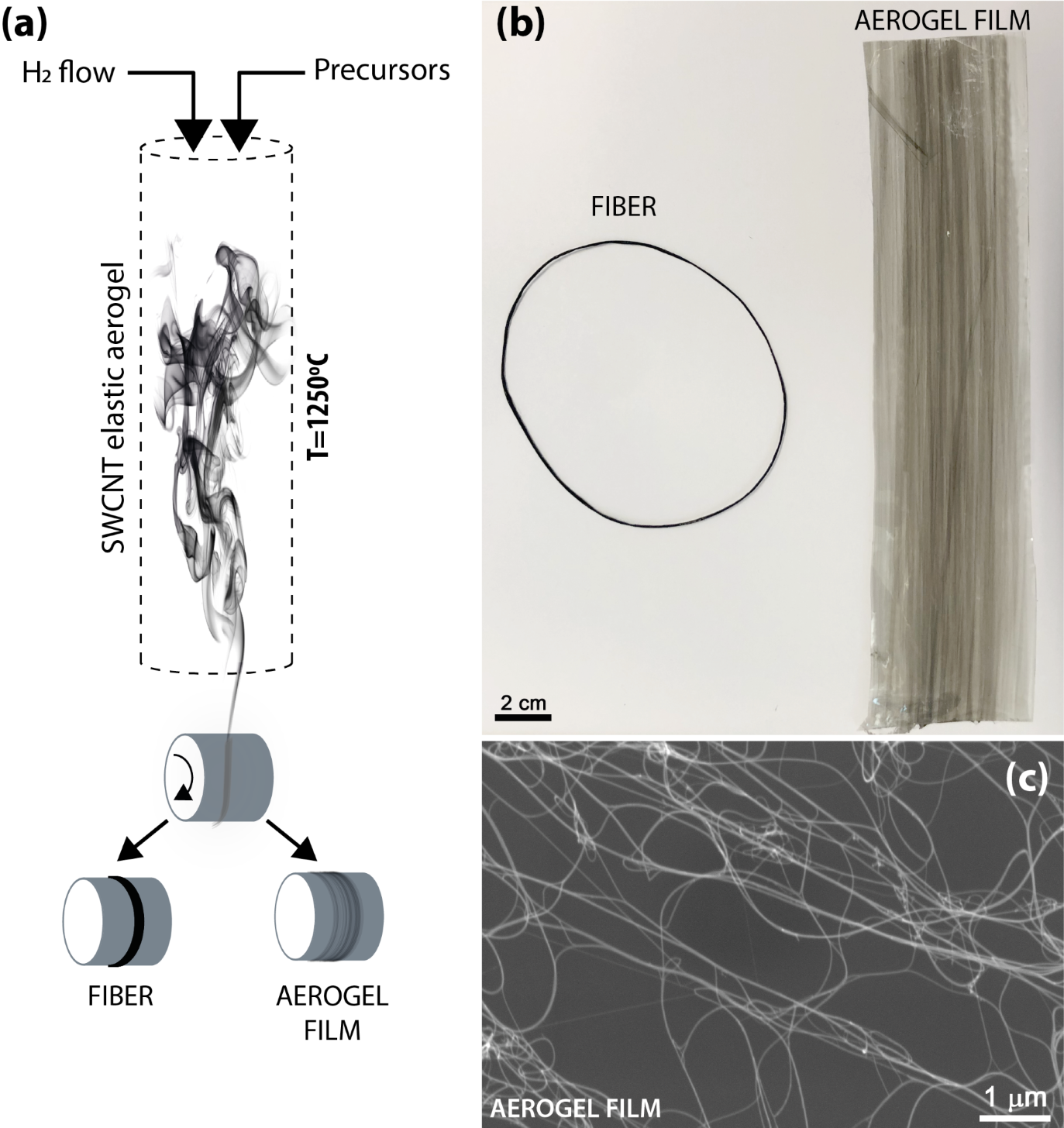}
  \caption{SWCNT aerogel samples used for molecular characterization and their correspondance to fibres (a) Schematic of the process to produce a densified fiber or an aerogel film by avoiding densification into a fiber. (b) Optical micrograph of a typical samples and (c) electron micrograph showing the the aerogel structure as a continous network of bundles}.
  \label{furnace}
\end{figure}

The strategy to study the molecular structure of SWCNT fibers consists in collecting a SWCNT aerogel at its exit from the FCCVD reactor (Fig.\ref{furnace}(a)). By avoiding the densification into a SWCNT fiber, we retain the open structure of the aerogel, thus providing access to individualized bundles and other structural features that are inaccessible in a fully densified fiber. The resulting sample is similar to a transparent conductive film\cite{yu_recent_2016} (Fig.\ref{furnace}(b)). The SWCNT bundles form a continuous network by virtue of the exceptionally long length of the constituent SWCNTs (1mm), but form an open structure whose projected area covers less than 35 \% of the substrate (Fig. \ref{furnace}(c) and ESI). 

\begin{figure}[hbt!]
\centering
  \includegraphics[width=6.5cm]{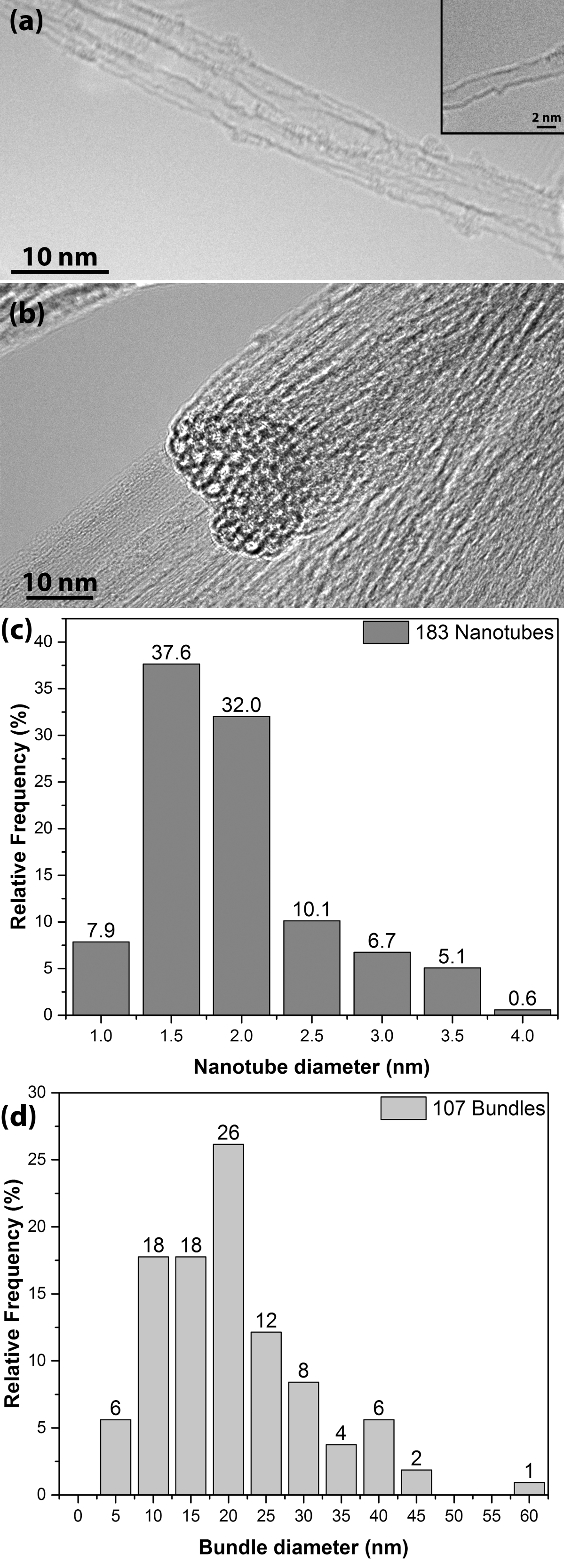}
  \caption{SWCNTs and bundles observed in aerogels produced by FCCVD: (a-b) Typical HRTEM images showing SWCNT and their association into bundles and (c-d) statistical analysis of SWCNTs and bundle diameters obtained from over 100 HRTEM images.}
  \label{TEM}
\end{figure}

A general overview of the morphology of the bundle network and constituent SWCNTs is gained by analysis of the aerogel directly deposited on a TEM grid. We obtain full diameter distributions of the bundles and individual SWCNTs from analysis of over $100$ HRTEM micrographs. Some examples of micrographs are included in Figure \ref{TEM}. As expected, the SWCNT are of relatively large diameter ($>$ 1 nm) (Fig.\ref{TEM}(c)), at least compared to those produced by standard substrate-based chemical vapor deposition (CVD) in the absense of promotors, but not sufficiently large to collapse into ribbons. The bundle diameter distribution observed by TEM spans from 5-60 nm and peaks at around 20 nm (Fig.\ref{TEM}(d)), which is comparable to the average bundle lateral size determined by small-angle X-ray scattering (9 nm) \cite{Cleis}. 

\begin{figure}[h]
\centering
  \includegraphics[width=8 cm]{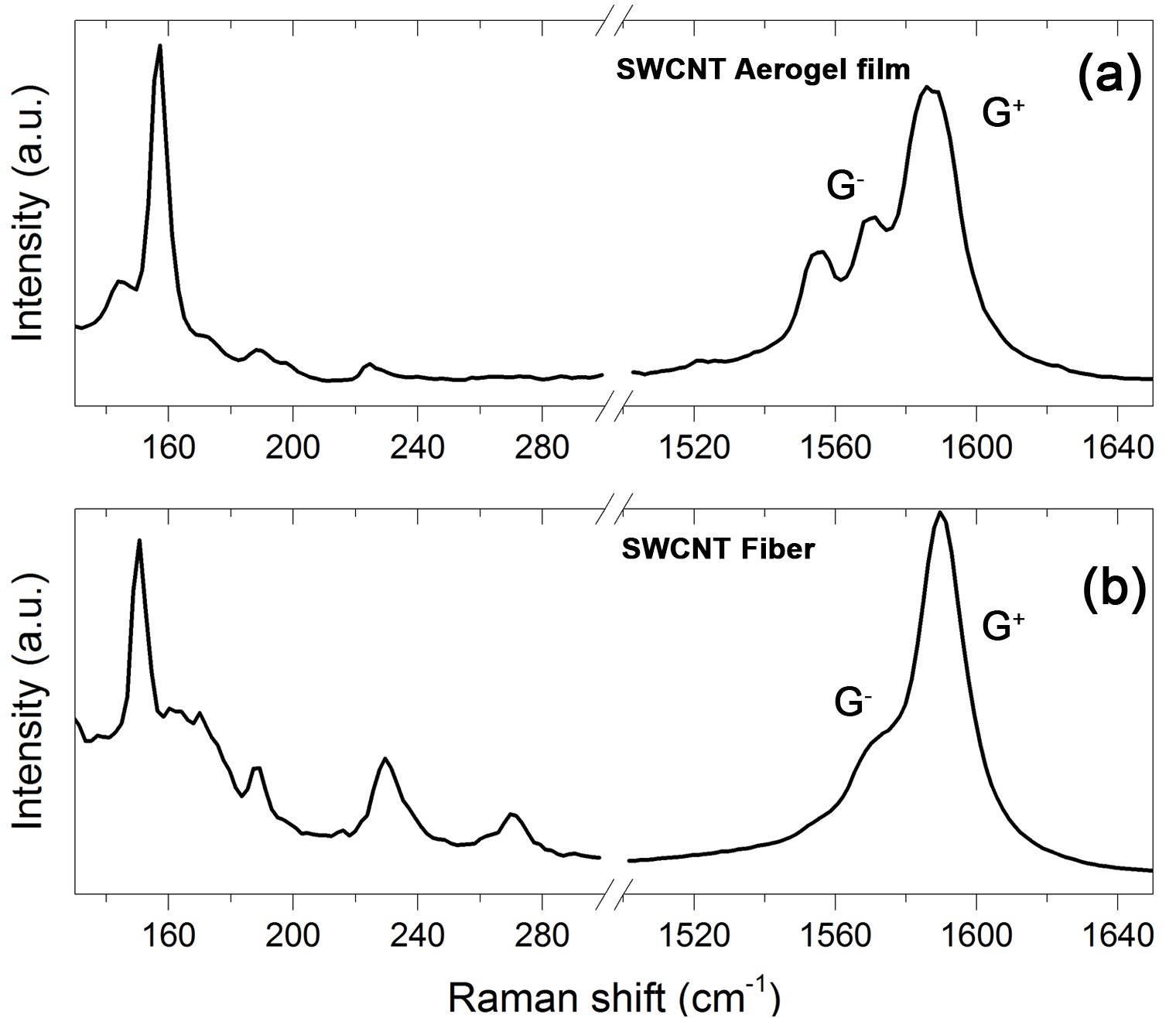}
  \caption{Emergence of fine Raman features in SWCNTs aerogels compared with densified fibers. (a) Raman spectrum showing semiconducting componets in the G band and well resolved RBM peaks, compared to (b) the spectrum for a densified fiber of the same composition showing only a metallic G$^{-}$ band lineshape. } 
  \label{spectra_comp}
\end{figure}

A comparison of Raman spectra for the SWCNT aerogel film and SWCNT fiber (Fig. \ref{spectra_comp}(a-b)) shows the presence of RBM peaks at the lower frequency range (100-300 cm$^{-1}$) and the highly graphitic character of the bundles manifested by a low I$_{D}$/I$_{G}$ ratio of 0.06$\pm$0.02. But there are large differences between spectra. The most striking is the presence of the two well-resolved components of the low frequency G$^{-}$ band in the SWCNT aerogel film, which originate from semiconducting nanotubes, to our knowledge not previously observed in SWCNT fibers. In contrast, the characteristic G band of the macroscopic SWCNT fiber consistently presents \textit{only} the G$^{-}$ band related to the metallic nanotubes, as reported in similar SWCNT fibers \cite{sundaram_continuous_2011,barnard_role_2016, aleman_inherent_2016} and related SWCNT bundle materials \cite{liu_diameter-selective_2008,ding_highly_2017}. 

\subsection{Resolving metallic and semiconducting SWCNT bundles} \label{met_sc}

\begin{figure}[hbt!]
\centering
  \includegraphics[width=8 cm]{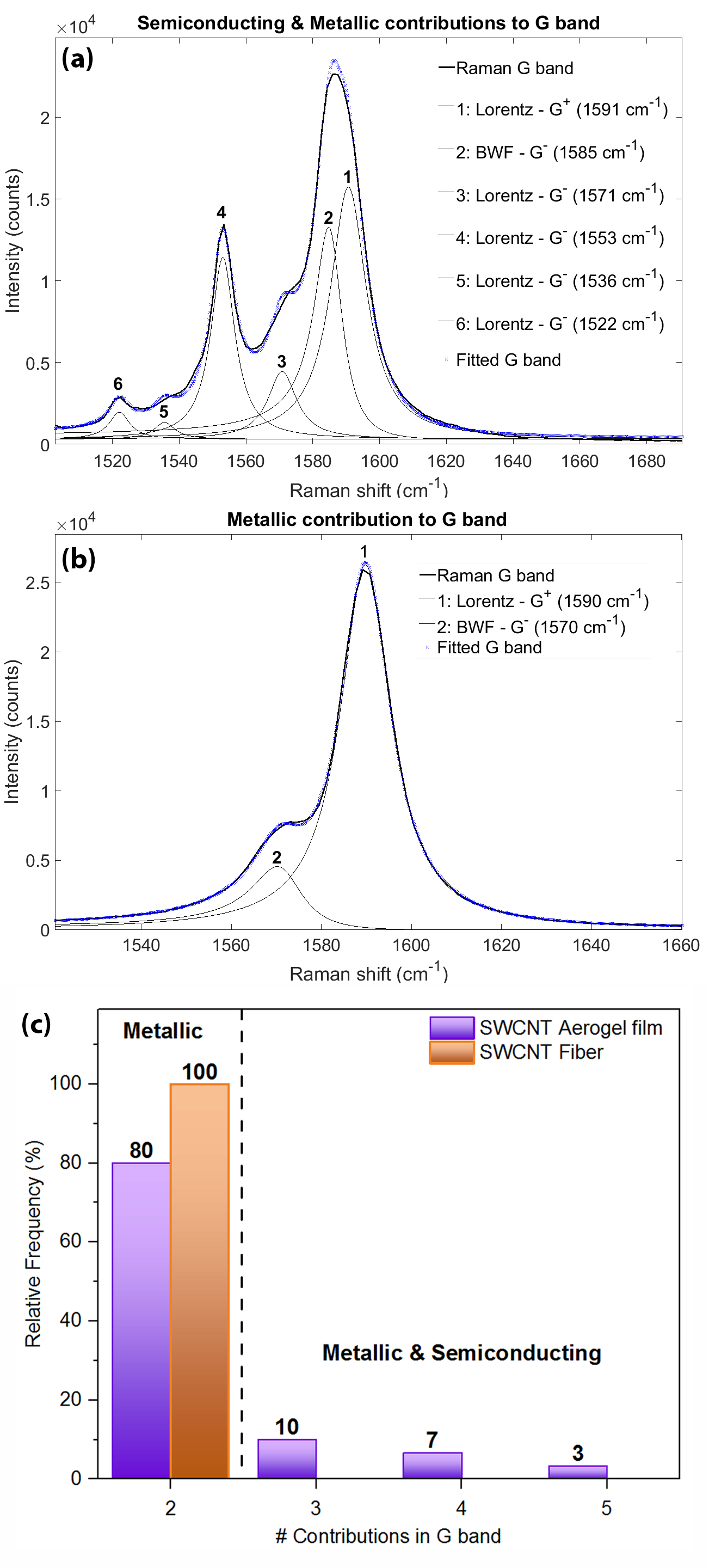}
  \caption{Raman G peaks for $\lambda$=532 nm obtained from individualized bundles in the SWCNT aerogel film. (a) Well-resolved Lorentzian G$^{-}$ components together with metallic G$^{-}$ contribution in a predominantly semiconducting bundle. (b) G$^{-}$ peak with BWF lineshape showing predominance of M-SWCNTs in the bundle probed.}
  \label{g_band}
\end{figure}

 A closer look at the G band lineshape provides further details about the presence of semiconducting and metallic SWCNTs: in semiconducting (SC) SWCNTs the G$^{-}$ band has  Lorentzian lineshape, whereas in metallic (M) SWCNTs it has  Breit-Wigner-Fano (BWF) lineshape. Fig.\ref{g_band}(a) shows and example of a bundle that clearly exhibits intense semiconducting features with Lorentzian lineshapes at 1571, 1553, 1536 and 1522 cm$^{-1}$, as well as contribution from metallic SWCNTs at 1585 cm$^{-1}$ (more spectra showing SC-SWCNT contributions in ESI). This is indicative of a bundle of mainly SC-SWCNTs. Bundles of M-SWCNTs are also present in the SWCNT aerogel, as in Fig. \ref{g_band}(b), with a predominant BWF G$^{-}$ band caused by a gapless plasmon mode (0.2 eV) along the axial CNT direction, which is enhanced in nanotube bundles through the formation of a plasmon band \cite{kempa_gapless_2002} due to coulomb interactions between CNTs\cite{lin_-plasmons_1998}. The plasmon band position depends on the SWCNT diameter \cite{jorio_gband_2002}, hence the relatively wide range observed for the SWCNT aerogel. In commercial samples of sorted M-SWCNTs the intensity ratio between G$^{-}_{BWF}$ and G$^{+}$ bands strongly increases with bundle thickness \cite{jiang_strong_2002,paillet_vanishing_2005}, reaching values of I$_{G^{-}_{BWF}}$/I$_{G^{+}}>1$ for bundles as small as 6nm. The presence of semiconducting nanotubes in a bundle of predominantly M-SWCNTs produces a reduction of this ratio. Indeed, in the SWCNT aerogels analyzed in this work, we find that even bundles without Lorentzian $G^{-}$ peaks, the ratio I$_{G^{-}_{BWF}}$/I$_{G^{+}}\leq0.5$ which is an indirect proof of the presence of SC-SWNCTs. 
 But as a rapid screening method, the most direct evidence of  SC-SWCNTs in the aerogel film is the presence of more than two peaks in the G band region: G$^{+}$, collective metallic (BWF) G$^{-}$ and at least one semiconducting (Lorentzian) G$^{-}$. Using this criteria, we obtain an abundance of metallic and semiconducting SWCNTs throughout the aerogel film, presented in Fig. \ref{g_band}(c). The abundance of semiconducting features in the aerogel contrasts with their apparent absence in the SWCNT fibers made up of the same building blocks. Overall, our analysis of around 30 spectra indicates that semiconducting SWCNTs are present in 20\% of the bundles probed.    
 

\subsection{SWCNT assignation} \label{families}

Retaining the bundle structure of SWCNT fibers by collapsing them into an open network enables an analysis of SWCNT chiral angle distribution through better resolved RBM peaks, and more importantly, their correspondence with the G$^{-}$ lineshape of the bundle. We assign RBM modes to \textit{(n,m)} indices by means of SWCNT family branches \textit{(2n+m=constant)} and using the procedure developed by Maultzsch and co-workers \cite{maultzsch_radial_2005}. The method is based on the relationship between optical transitions and SWCNT diameter, represented in the Kataura plot \cite{kataura_optical_1999-1}, taking into account that characteristic branches of families of SWCNTs  show a downshift in transition energies arising from Van der Waals interactions.

\begin{figure}[hbt!]
\centering
  \includegraphics[width=8.5cm]{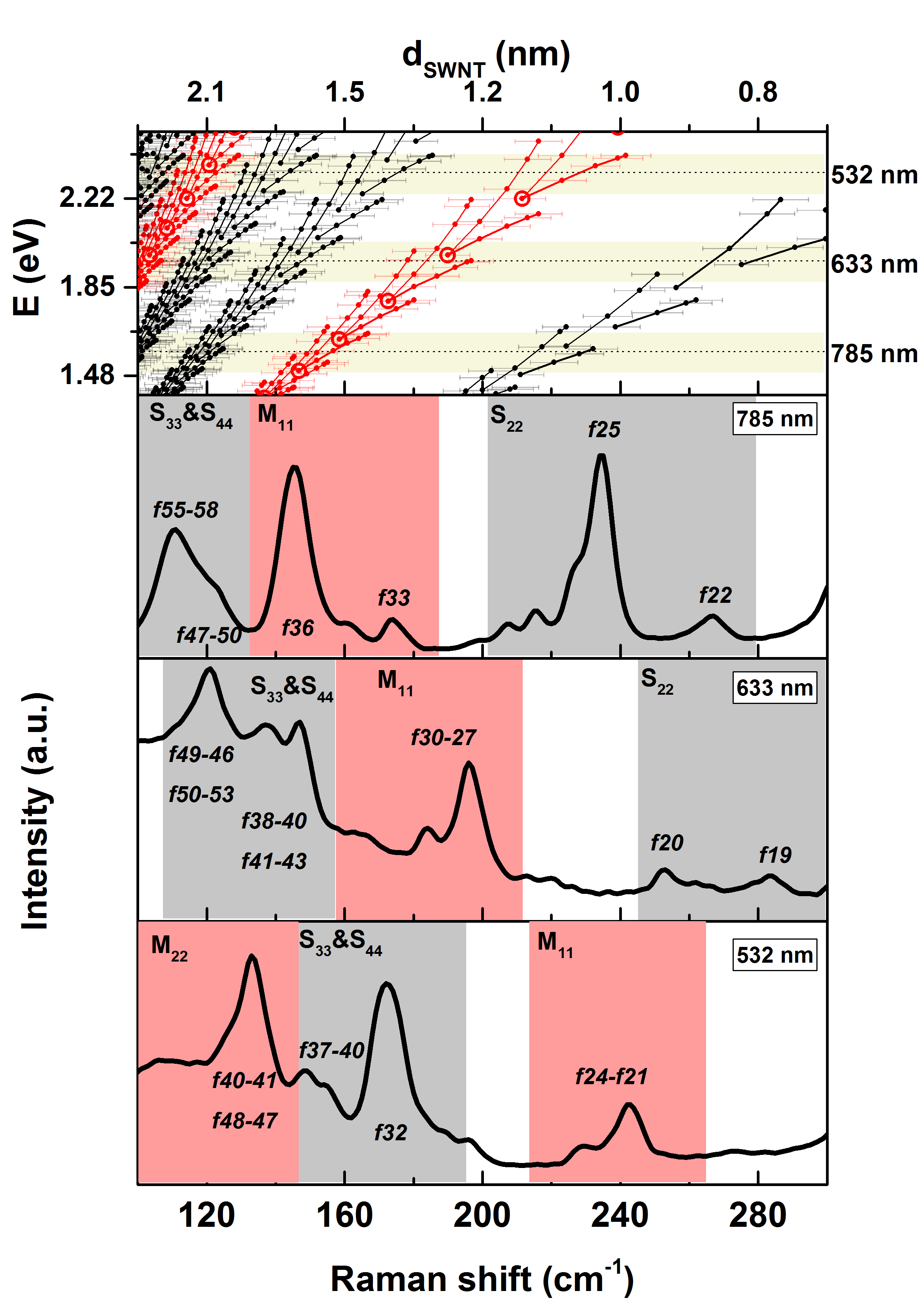}
  \caption{Kataura plot of transition energies vs inverse diameter for metallic (black) and semiconducting (red) SWCNTs and RBMs of a bundle for 785, 633 and 532 nm laser wavelengths with the indexed metallic or semiconducting branches (M$_{ii}$ or SM$_{ii}$) as well as the family assignation.}
  \label{assignation}
\end{figure}

Figure \ref{assignation} shows examples of the most common RBM peaks observed in the SWCNT aerogel film probed with three different energy lasers (785, 633 and 532 nm). Superimposing the Kataura plot enables identification of the SWCNT families (\textit{f}). Key for this step though, is the metallicity character observed in the G$^{-}$ lineshape, as it enables the accurate location of the spectra onto the Kataura plot. Only by analizing separated bundles (instead of the dense fiber) are SC-SWCNTs exposed and thus the family assignation can be made. Using a 785 nm laser, for example, there are strong RBMs at 140 and 240 $cm^{-1}$; together with evidence of mixed BWF and Lorentzian G$^{-}$, they can thus be assigned to transitions M$_{11}$ of \textit{f36} and S$_{22}$ of \textit{f25}, respectively. Further assignation of RMBs from larger-diameter SWCNTs that are less separated in the Kataura plot can then proceed. A summary of families observed are included in ESI. 

\begin{figure}[h]
\centering
  \includegraphics[width=8.5cm]{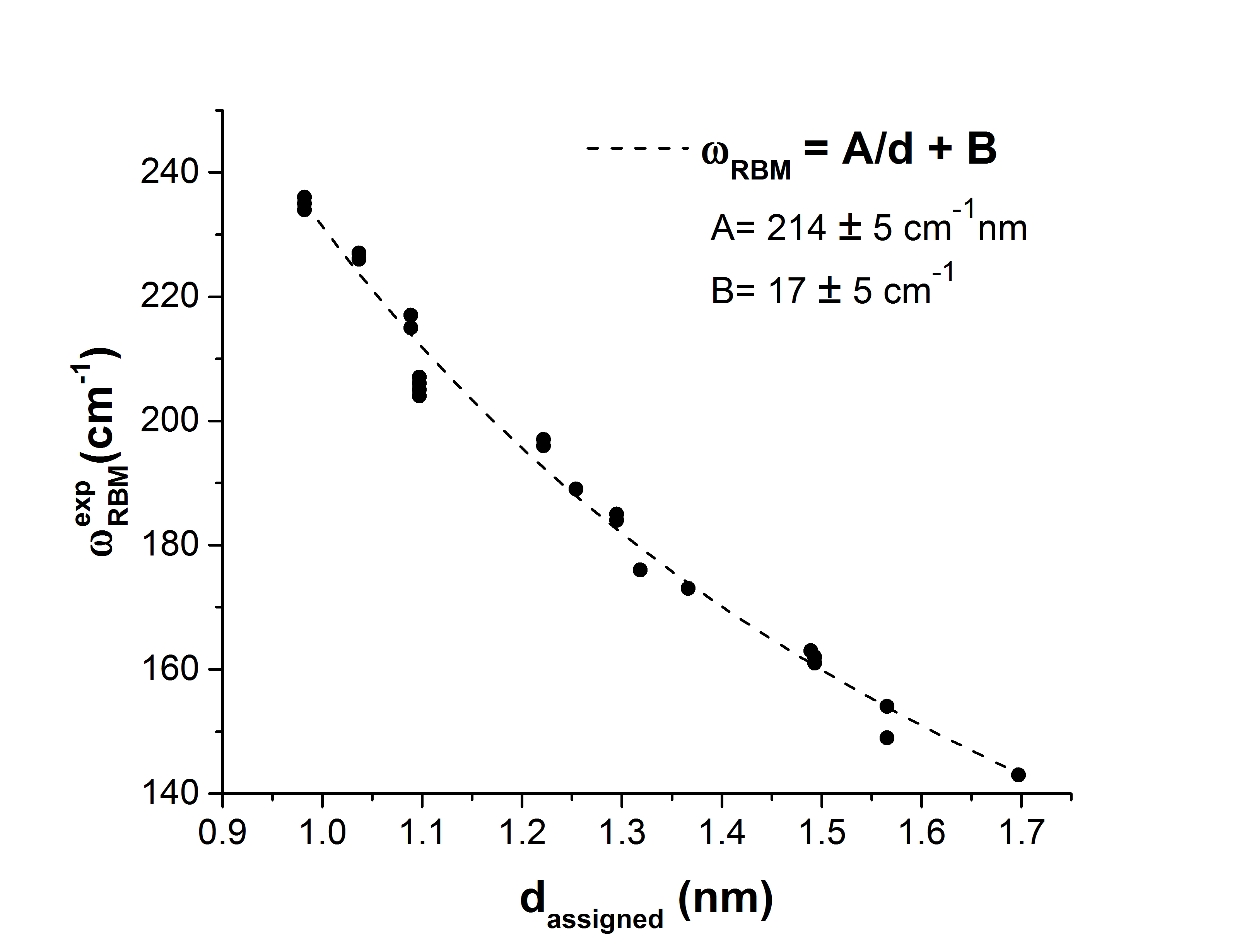}
  \caption{Graphic representation of the experimental RBM Raman shifts and their corresponding assigned nanotube diameter.}
  \label{wd}
\end{figure}

Next, after identification of families on the Kataura plot, specific SWCNTs can be assigned to RBM peaks by identifying those whose optical transition fall within $\pm$50 meV of the laser excitation energy, and are thus in resonance. For the spectra using a 785 nm laser, for example, the most likely SWCNT of \textit{f25} with an RBMs at $\sim$236 cm$^{-1}$ is the semiconducting (12,1) with a diameter of 0.98 nm. The assignation of all observed RBMs in the sample is included in ESI. Its accuracy is confirmed by the expected reciprocal dependence between RBM Raman shift and SWCNT diameter obtained experimentally (Figure \ref{wd}), leading to the relationship 

\begin{equation}
  \omega_{RBM}(cm^{-1})=\dfrac{214}{d}+17
  \label{eq}
 \footnotesize
\end{equation}
\footnotesize
\noindent
\footnotesize
The values for the coefficients A= (214$\pm$5) cm$^{-1}$·nm and B= (17$\pm$5) cm$^{-1}$ are in the range reported for individualized, surfactant-wrapped SWCNTs in aqueous suspensions \cite{maultzsch_radial_2005,fantini_optical_2004}, as well as for aligned SWCNTS grown on a Si wafer and including both individualized and bundled tubes \cite{araujo_third_2007}.

\begin{figure*}[h]
\centering
  \includegraphics[width=15 cm]{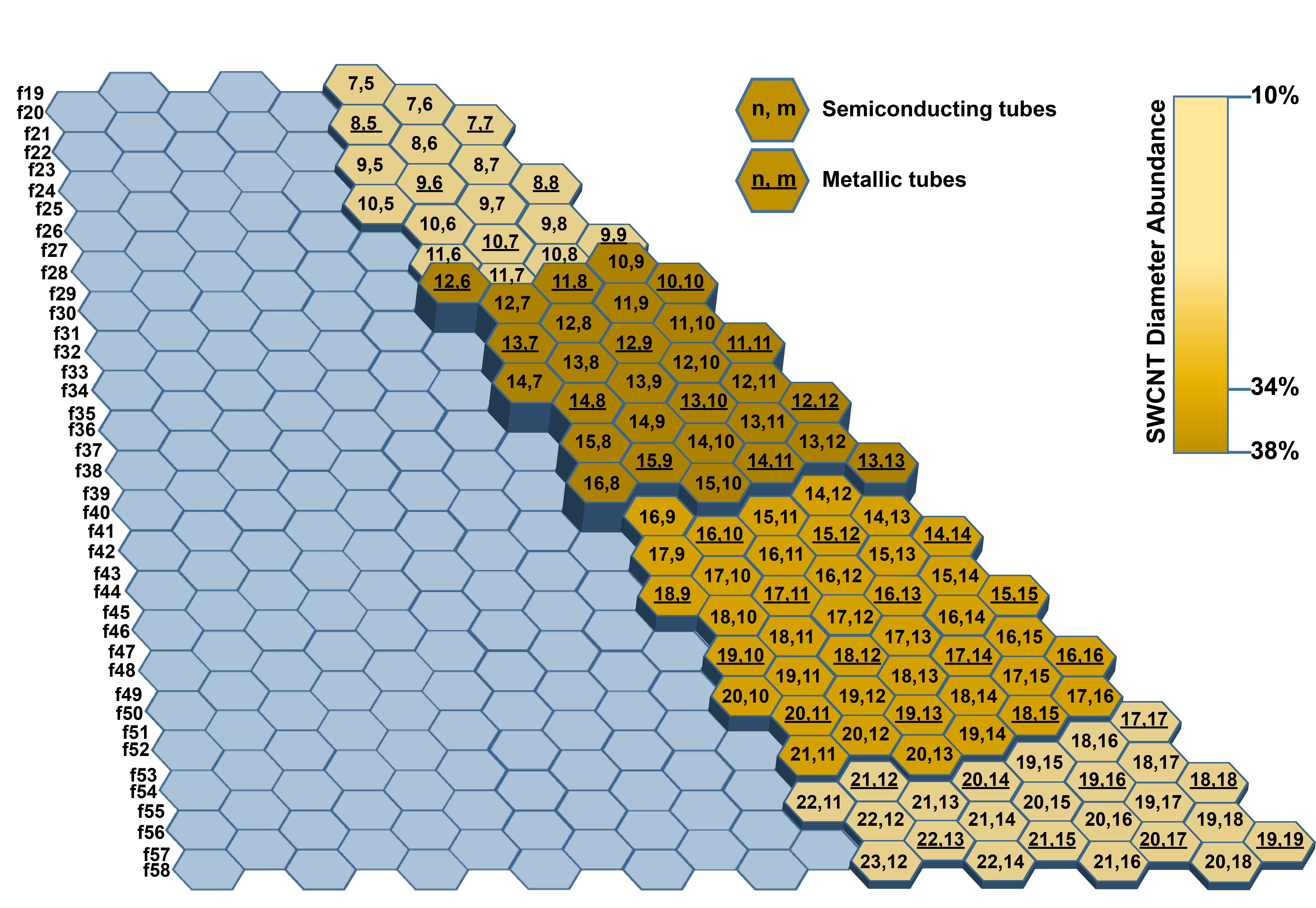}
  \caption{Distribution of chiral indices \text{\textit{(n,m)}} for SWCNTs in aerogels produced by FCCVD, determined by combining HRTEM diameter measurements, chiral angle determination by electron diffraction, and confirmed by Raman spectroscopy measurements over large areas of SWCNT aerogel samples.}
  \label{map}
\end{figure*}

Overall, the results after full assignation show compelling evidence of the presence of semiconducting nanotube families together with the metallic ones, and which emerge in the open network aerogel by avoiding masking of semiconducting features in the spectra by metallic SWCNTs. Analysed in this manner, the data from Raman spectroscopy is in full agreement with TEM observations. Figure \ref{map} shows a map of \textit{(n,m)} indices based on the nanotube diameter data extracted from the HRTEM images and the chiral angle distribution previously determined from electron diffraction measurements \cite{aleman_inherent_2016}. The families of SWCNTs determined by Raman from measurements on the open network and accessible with the three laser lines used in the present study, are also marked on the map. The agreement between them is clear, providing a consistent interpretation of the molecular structure of SWCNT fibers obtained using the two characterization techniques.

Finally, we highlight some challenges for SWCNT molecular control that stem from the large diameter and wide distribution observed in SWCNT fibers produced by FCCVD. The most immediate one is the increasing difficulty for \textit{(n,m)} indices assignation due to the overlap in possible CNT families when the measurements are conducted with standard 3-4 laser lines. Moreover, for the predominant diameters observed in these SWCNT aerogels (1.5 - 2 nm), the corresponding energy transitions are of third and fourth order, S$_{33}$ and S$_{44}$, and thus more likely to present energy deviations from Kataura plot due to larger delocalization of electrons in higher lying levels compared with lower order transitions \cite{araujo_third_2007}. This characterization challenge can be partially solved using a tunable laser Raman spectrometer to broaden the selection of incident energies to resolve the optical transitions of a larger number of SWCNTs types. However the ultimate benefit of synthesizing a narrower distribution of smaller-diameter SWCNTs is the reduction in possible \textit{(n,m)} indices. The number of geometrically possible chiral indices for the diameter range of 0.75 - 2.25 nm observed by HRTEM in the samples in this study, is $\sim$240. Even if the chiral angle distribution is biased towards armchair (20$^\circ$-30$^\circ$) \cite{aleman_inherent_2016}, this landscape appears too broad at present, as illustrated in the comparison with conventional substrate-grown SWCNTs in ESI.

Future work aimed at molecular control will requires a chain of characterisation methods of increasing resolution and decreasing probe size, the first of which would be the Raman technique introduced here. We envisage the characterization protocol to follow this sequence: i)  determination of SWCNT family distribution by Raman using 3-4 laser lines (resolution of $\pm$50 meV); if the distribution is narrow, for example, less than $\approx 3$ families observed per laser energy, then ii) perform HRTEM electron diffraction and diameter measurements (requiring resolving diameter and chiral angle differences of $<0.1$ nm  and 1$^\circ$ respectively) and iii) carry out absorption and emission measurements (example of optical absorption spectrum for an aerogel ESI). An interesting possibility is to combine these tools with high-throughput synthesis and analysis methods \cite{nikolaev_autonomy_2016}.

\section{Conclusions}

This work presents a study on the molecular structure of SWCNTs in macroscopic fibers produced by FCCVD. By retaining the open structure of the aerogel directly produced during the synthesis stage, we produce samples consisting of a continuous open network of bundles sufficiently separated to be analyzed individually by HRTEM and Raman spectroscopy. Avoiding aggregation of bundles in the aerogel unveils the abundant presence of SC-SWCNTs in these fibers, manifested clearly by a Lorentzian lineshape of the low frequency G$^{-}$ bands. In the fully densified SWCNT fiber or other aggregated ensembles, these SC-SWCNTs are masked by M-SWCNTs, resulting in a BWF lineshape due to the formation of a plasmon band.  
Equipped with a clearly resolved G$^{-}$ band, a full map of SWCNT families and chiral indices can be produced from assignment of RBMs to optical transitions according to the Kataura plot. The empirical relationship found $\omega_{RBM}$=214/d+17 indicates good agreement in the assignation, and gives empirical coefficients in the range observed for SWCNTs with strong Van der Waals interaction with the medium, as is expected for bundles. The SWCNTs observed with the three laser lines employed, is in agreement with the distribution obtained by combining HRTEM diameter measurements and chiral angle determination from electon diffraction.  
The method presented offers a tool to characterize the molecular structure of SWCNT fibers (in the aerogel state) probing much larger areas and much faster than alternative techniques. Future work is directed at exploiting it for rapid SWCNT fiber optimization, and more importantly, to establish clear correlations between constituent SWCNTs and bulk properties. The open structure format would lend itself, for example, to transport measurements or surface potential characterisation methods that could resolve the conduction mechanism in bundles with different constituent SWCNTs.  

\section*{Acknowledgements}

The authors are grateful to Professor E. Pérez for access to Raman spectrometers, and acknowledge generous financial support from the European Union Seventh Framework Program under grant agreement 678565 (ERC-STEM) and from the Air Force Office of Scientific Research of the US (NANOYARN FA9550-18-1-7016). 

\section*{Appendix A. Supplementary data}

Supplementary data to this article can be found online at .... 



\section*{References}

\bibliography{SWRaman.bib}

\end{document}